\documentclass[12pt]{article}
\def\be{\begin{equation}}
\def\ee{\end{equation}}
\def\bea{\begin{eqnarray}}
\def\eea{\end{eqnarray}}
\input epsf.tex

\catcode`@=12
\begin{document}
\thispagestyle{empty}
\begin{flushright}
UCRHEP-T406\\
March 2006\
\end{flushright}
\vspace{0.5in}
\begin{center}
{\LARGE \bf Leptogenesis Below the\\ Davidson-Ibarra Bound\\}
\vspace{1.0in}
{\bf Ernest Ma\\}
\vspace{0.2in}
{\sl Physics Department, University of California, Riverside,
California 92521, USA\\}
\vspace{0.5in}
{\bf Narendra Sahu and Utpal Sarkar\\}
\vspace{0.2in}
{\sl Physical Research Laboratory, Ahmedabad 380009, India\\}
\vspace{1.0in}
\end{center}
\begin{abstract}\
The observed baryon asymmetry of the Universe is suitably created in thermal
leptogenesis through the out-of-equilibrium decay of $N_1$, the lightest
of the three heavy singlet neutral fermions which anchor the seesaw 
mechanism to obtain small Majorana neutrino masses.  However, this
scenario suffers from the incompatibility of a generic lower bound on
the mass of $N_1$ and the upper bound on the reheating temperature
of the Universe after inflation. A modest resolution of this conundrum
is proposed.
\end{abstract}
                                                                                
\newpage
\baselineskip 24pt
The canonical seesaw mechanism \cite{seesaw} for small Majorana neutrino
masses requires the existence of three heavy singlet neutral fermions $N_i$
so that
\begin{equation}
{\cal M}_\nu = - {\cal M}_D {\cal M}_N^{-1} {\cal M}_D^T\,
\label{neu_mass}
\end{equation}
where ${\cal M}_D$ is the $3 \times 3$ Dirac mass matrix linking
$\nu_\alpha$ with $N_i$ through the Yukawa interactions $h_{\alpha i}
(\nu_\alpha \phi^0 -l_\alpha \phi^+) N_i$, where $(\phi^+,\phi^0)$ is
the Higgs doublet of the Standard Model (SM). In the early Universe, a
lepton asymmetry may be generated \cite{fy86} by the out-of-equilibrium
decay of the lightest $N_i$ (call it $N_1$), which gets converted into a
baryon asymmetry
through the interactions of the SM sphalerons \cite{kr85} which conserve
$B-L$, but violate $B+L$, where $B$ and $L$ are baryon and lepton number
respectively. The existence of $N_i$ explains thus at the same time why
both neutrino masses as well as the observed baryon asymmetry of the Universe
(BAU) are nonzero and small.

In the context of cosmology, the Universe goes through a period of
inflation and then gets reheated to a certain maximum temperature
$T_h$ which is limited by the possible overproduction of gravitinos
\cite{gravitino}, if the underlying theory of matter is supersymmetric.
It has been shown \cite{di02} that the generic lower bound on the 
mass $M_1$ of $N_1$
for successful thermal leptogenesis is dangerously close to being higher
than $T_h$. This means that $N_1$ is not likely to be
produced in enough abundance to generate the BAU.

To avoid this problem, several ideas have been discussed in the 
literature~\cite{ideas}. In particular, a more recent proposal is 
to consider the flavour issues in thermal leptogenesis~\cite{riotoetal}. 
Here a new and very simple solution is proposed. In addition to the three
$N_i$ ($i=1,2,3$), we add one more singlet fermion $S$ together with a
discrete $Z_2$ symmetry, under which $S$ is odd and all other fields are
even.  This $Z_2$ symmetry prevents the Yukawa coupling of $S$ to the
usual lepton and Higgs doublets, but is allowed to be broken softly and
explicitly by the $N_i S$ terms.  Thus $S$ mixes with $N_i$ and its
(indirect) couplings to the leptons are naturally suppressed.
Specifically, the $7 \times 7$ mass matrix spanning
$(\nu_e,\nu_\mu,\nu_\tau,N_1,N_2,N_3,S)$ is given by
\begin{equation}
{\cal M} = \pmatrix{0 & 0 & 0 & m_{e1} & m_{e2} & m_{e3} & 0 \cr
0 & 0 & 0 & m_{\mu 1} & m_{\mu 2} & m_{\mu 3} & 0 \cr
0 & 0 & 0 & m_{\tau 1} & m_{\tau 2} & m_{\tau 3} & 0 \cr
m_{e1} & m_{\mu 1} & m_{\tau 1} & M_1 & 0 & 0 & d_1 \cr
m_{e2} & m_{\mu 2} & m_{\tau 2} & 0 & M_2 & 0 & d_2 \cr
m_{e3} & m_{\mu 3} & m_{\tau 3} & 0 & 0 & M_3 & d_3 \cr
0 & 0 & 0 & d_1 & d_2 & d_3 & M_S}.
\label{mass-matrix}
\end{equation}

In canonical leptogenesis without $S$, the lightest right-handed neutirno
$N_1$ decays into either $l^- \phi^+$ and $\nu \phi^0$, or $l^+ \phi^-$ and
$\bar \nu \bar \phi^0$. Thus a CP asymmetry may be established from the
interference of the tree-level amplitudes with the one-loop vertex and
self-energy corrections~\cite{fps}. The decay rate
\begin{equation}
\Gamma_1 = {(h^\dagger h)_{11} \over 8\pi} M_1
\end{equation}
is compared against the expansion rate of the Universe
\begin{equation}
H(T) = 1.66 g_*^{1/2} {T^2 \over M_{Planck}}
\label{hubble}
\end{equation}
at $T \sim M_1$, where $g_* \simeq 230$ is the effective number of
relativistic degrees of freedom in the Minimal Supersymmetric Standard
Model (MSSM) and $M_{Planck} = 1.2 \times 10^{19}$ GeV. This means
that a \underline{lower} bound on $M_1$ may be established by first
considering the out-of-equilibrium condition
\begin{equation}
H(T=M_1) > \Gamma_1.
\label{out-of-equim}
\end{equation}
Let $K_1=\Gamma_1/H(T=M_1)$, then the above condition requires $K_1<1$,
but even if $K_1>1$, a reduced lepton asymmetry may still be generated,
depending on the details of the Boltzmann equations which quantify the
deviation from equilibrium of the process in question.

The baryon-to-photon ratio of number densities has been measured
\cite{eta_b} with precision, i.e.
\begin{equation}
\eta_B \equiv {n_B \over n_\gamma} = 6.1^{+0.3}_{-0.2}\times 10^{-10}.
\label{b-asym}
\end{equation}
In canonical leptogenesis, the thermal 
production of $N_1$ in the early Universe after reheating implies~\cite{bp99}
\begin{equation}
\eta_B \sim {\epsilon_1\over 10 g_*}\,,
\end{equation}
where a typical washout factor of 10 has been inserted, and the CP
asymmetry $\epsilon_1$ is given by
\begin{equation}
\epsilon_1 \simeq -{3 \over 8\pi} \left( {M_1 \over M_2} \right)
{Im [(h^\dagger h)^2]_{12} \over (h^\dagger h)_{11}}\,,
\label{epsilon_1}
\end{equation}
assuming $M_1 << M_2 << M_3$. To get the correct value of $\eta_B$, 
$\epsilon_1$ should be of order $10^{-6}$. However, using the 
Davidson-Ibarra upper bound on the CP asymmetry~\cite{di02}
\be
|\epsilon_1|\leq \frac{3M_1}{8\pi v^2}\sqrt{\Delta m_{atm}^2}.
\label{cp-bound}
\ee
this would imply $M_1 > 4 \times 10^{9}$ GeV.  Since $T_h$ is not likely 
to exceed $10^9$ GeV, this poses a problem for canonical leptogenesis. 

In the present model, the addition of $S$ allows the choice of $M_S < M_1$.
Since the mixing of $S$ with $N_i$ comes from the breaking of the assumed
$Z_2$ symmetry of the complete Lagrangian, the parameters $d_i$ are naturally
small compared to $M_i$.  The induced couplings of $S$ to leptons are
suppressed by factors of $d_i/M_i$ compared to those of $N_i$, with its 
decay rate given by
\begin{equation}
\Gamma_S = \sum_i {(h^\dagger h)_{ii} \over 8 \pi}
\left( {d_i \over M_i} \right)^2 M_S\,.
\label{gamma_S}
\end{equation}
The condition for the departure from equilibrium, i.e. 
Eq.~(\ref{out-of-equim}), during the decays of $S$ can then be 
satisfied at a much lower mass.

\begin{figure}[!thb]
\vskip 1.25in
\epsfxsize=120mm
\hskip -.95in
\centerline{\epsfbox{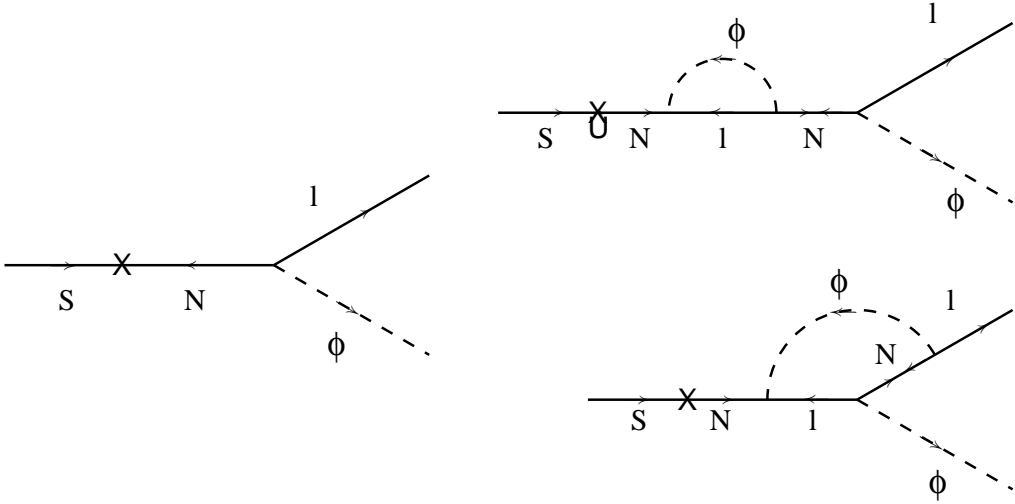}}
\vskip -.5in
\caption{Tree-level and one-loop (self-energy and
vertex) diagrams for $S$
decay, which interfere to generate a lepton asymmetry.  }
\label{lepfg}
\end{figure}

The CP asymmetry generated by the decays of $S$ comes from the interference
of the tree-level and one-loop diagrams of Figure \ref{lepfg}. Consider
the case where $S$ mixes only with $N_1$.  Both the numerator and denominator
of Eq. (\ref{epsilon_1}) are then suppressed by the same $(d_1/M_1)^2$ 
factor, and we obtain
\begin{equation}
\epsilon_{S_1} \simeq -{3 \over 8\pi} \left( {M_S \over M_2} \right)
{Im [(h^\dagger h)^2]_{12} \over (h^\dagger h)_{11}}\,.
\label{epsilon_s}
\end{equation}
Comparing Eq.~(\ref{epsilon_s}) with Eq.~(\ref{epsilon_1}), we see 
that we have not gained anything because $M_S$ is subject to the same 
lower bound as $M_1$ through Eq.~(\ref{cp-bound}).  Of course, we can 
adjust $(d_1/M_1)$ to make $K_S<1$ to avoid any washout, but then 
the thermal production of $S$ through its inverse decay (which 
prefers $K_S>1$) will be suppressed and we have again the typical 
reduction by about a factor of ten, as shown in Ref.~\cite{bp99}.

To reduce $M_S$ below the Davidson-Ibarra bound of $4 \times 10^9$ 
GeV, we may consider the case where $S$ mixes only with $N_2$, then
\begin{equation}
\epsilon_{S_2} \simeq -{3 \over 8\pi} \left( {M_S \over M_1} \right)
{Im [(h^\dagger h)^2]_{21} \over (h^\dagger h)_{22}}\,.
\label{epsilon_s2}
\end{equation}
This would allow $M_S$ to be smaller by a large factor, as shown below.
Consider the following approximate Yukawa matrix
\begin{equation}
h_{\alpha i} \simeq \pmatrix{0 & 0 & 0 \cr h_1 & -i h_2 & 0 \cr h_1 &
-i h_2 & 0}\,,
\end{equation}
where $h_{1,2}$ are real, from which we obtain 
\begin{equation}
(h^\dagger h)_{22} \simeq 2 h_2^2, ~~~ [(h^\dagger h)^2]_{21} \simeq
4i h_1 h_2 (h_1^2 + h_2^2)\,.
\label{model_const}
\end{equation}
As for the neutrino mass matrix, it is given by
\begin{equation}
{\cal M}_\nu = h_{\alpha i} M_i^{-1} h_{\beta i} v^2 =
m_0 \pmatrix{0 & 0 & 0 \cr 0 & 1 & 1 \cr 0 & 1 & 1}\,,
\end{equation}
where $m_0/v^2 = h_1^2 M_1^{-1} - h_2^2 M_2^{-1}$. This neutrino mass
matrix has maximal $\nu_\mu-\nu_\tau$ mixing and $m_3 \simeq
\sqrt{\Delta m^2_{atm}}$ and $m_{1,2} \simeq 0$, which is a reasonable
approximation of the present data on neutrino oscillations. Since $M_2$
is assumed much greater than $M_1$, we have thus
\begin{equation}
m_3 \simeq 2 h_1^2 v^2/M_1\,.
\end{equation}
Now Eq.~(\ref{epsilon_s2}) can be expressed as
\begin{equation}
\epsilon_{S_2} \simeq - {3 M_S m_3 \over 8 \pi v^2} \left( {h_1^2
+h_2^2 \over h_1 h_2}\right).
\end{equation}
Comparing this to the bound of Eq.~(\ref{cp-bound}), we see that 
$M_S$ may be lowered by the factor $h_1 h_2 / (h_1^2 + h_2^2)$. 
It may thus be reduced by, say a factor of 10 to $4 \times 10^8$ GeV, 
below the reheating temperature of $10^9$ GeV. 

In conclusion we have shown that the simple addition of an extra singlet 
to the usual three heavy neutral singlet fermions responsible for the 
seesaw mechanism in the MSSM offers a modest solution to the gravitino  
problem in canonical leptogenesis. 
                                                                            
%\newpage
This work was supported in part by the U.~S.~Department of Energy under
Grant No. DE-FG03-94ER40837.  EM thanks the Physical Research 
Laboratory, Ahmedabad, India for hospitality during a recent visit.
                                                                              
%\newpage
\bibliographystyle{unsrt}

\end{document}